\documentclass[letterpaper,prl,twocolumn,showpacs,superscriptaddress]{revtex4}
\usepackage{graphicx}
\usepackage{amsmath}
\usepackage{amssymb}

\begin{document}

\title{Quantum measurements of spatial conjugate variables:
Displacement and tilt of a Gaussian beam}

\author{V. Delaubert}
\affiliation{Laboratoire Kastler Brossel, UPMC, Case 74, 4 Place Jussieu, 75252 Paris cedex 05, France}
\author{N. Treps}
\affiliation{Laboratoire Kastler Brossel, UPMC, Case 74, 4 Place Jussieu, 75252 Paris cedex 05, France}
\author{C. C. Harb}
\affiliation{ARC Centre of Excellence for Quantum-Atom Optics, The
Australian National University, Canberra ACT 0200, Australia}
\author{Ping Koy Lam}
\affiliation{ARC Centre of Excellence for Quantum-Atom Optics, The
Australian National University, Canberra ACT 0200, Australia}
\author{Hans. A. Bachor}
\affiliation{ARC Centre of Excellence for Quantum-Atom Optics, The
Australian National University, Canberra ACT 0200, Australia}

\begin{abstract}
We consider the problem of measurement of optical transverse
profile parameters and their conjugate variable. Using multi-mode
analysis, we introduce the concept of detection noise-modes. For
Gaussian beams, displacement and tilt are a pair of transverse
profile conjugate variables.  We experimentally demonstrate their
optimal encoding and detection with a spatial homodyning scheme.
Using higher order spatial mode squeezing, we show the sub-shot
noise measurements for the displacement and tilt of a Gaussian
beam.
\end{abstract}

\pacs{42.50.-p; 42.50.Dv}

\maketitle

Quantum information protocols rely on the use of conjugated
variables of a physical system for information encoding.  For
single mode continuous variable systems, there are only a very
limited number of choices for such conjugate variable pairs.  For
example, phase and amplitude quadrature measurements with a
balanced homodyne detector, or polarization Stokes parameter
measurements using polarization discriminating detectors
\cite{BOWEN-KOROL}, are the common conjugate variables used
experimentally.

By not restricting ourselves to single mode analysis, we can use
the ability of a laser beam to transmit high multi-mode
information by extending these protocols to the transverse spatial
domain.  The transverse profile of the beam is then described by a
set of orthonormal modes that potentially allows a parallel
treatment of information.  Recently this parallel processing
scheme was used in single photon experiments to extend {\it
q-bits} to {\it q-dits} using modes with higher angular momentum
\cite{MAIR-LANGFORD}.

Such an improvement requires the perfect matching of the detection
system to the spatial information contained in the light beam.
Indeed, we have shown that a single detector extracts information from
only one specific transverse mode of the beam \cite{TREPS1}.  We call
this mode the {\it noise-mode of detection} since it is the only mode
contributing to the measurement noise.  As a consequence, information
encoded in any other mode orthogonal to the detection noise-mode is
undetected.  Moreover, noise-modes of detection are the transverse
spatial modes whose modulation in magnitude is transferred perfectly
to the detected output as a photocurrent.

The use of classical resources sets a lower bound to detection
performances, which is called the quantum noise limit (QNL) and
arises from the random time arrival of photons on the detector. In
the case of displacement measurement of a laser beam, the transverse displacement $d_{QNL}$ of a TEM$_{00}$ laser
beam corresponding to a signal to noise ratio of $1$, is given by 
$d_{QNL}=\frac{w_{0}}{2\sqrt{N}}$,
where $w_{0}$ is the waist of the beam, and $N$ its total number
of photons in the interval $\tau=1/RBW$, where $RBW$ is the resolution bandwidth \cite{treps2}. Note that the ability to resolve the signal relative to the noise can be further improved by averaging with the spectrum analyzer, by reducing the video bandwidth (VBW) and thus increasing the number of photons detected in the measurement interval, if the system has enough stability. For a $100~\mu m$ waist, $1~mW$ of power at a
wavelength of $\lambda=1~\mu m$, with $RBW=100~kHz$ and $VBW=100Hz$, the quantum noise limit is for instance given by $d_{QNL}=0.2~nm$, and the minimum measurable transverse displacement is $d_{min}=7~pm$.

In order to achieve a measurement sensitivity beyond the QNL, it
is a necessary and sufficient condition to fill the noise-mode of
detection with squeezed light \cite{TREPS1}.  As required by
commutation relations, a measurement of the conjugate variable
shows excess noise above the QNL.

In this paper, we first explain how spatial information can be encoded
onto a beam, and how optimized measurement of spatial properties of a
beam can be achieved classically.  As an example, we use the
displacement and tilt of a Gaussian laser beam \cite{HSU1} (which are
two spatial conjugate variables) to show the quantitative results of
signal-to-noise-ratio (SNR) measurements that surpass the quantum
noise limit.

Encoding information in the transverse plane of a laser beam can be
achieved by modulating any of its scalar parameters $p$ around a mean
value $p_{0}$.  This parameter can correspond to any deformation of
the transverse profile, such as displacement and tilt, which are
properties easy to visualize and to use in practice.  In the simple
case of a TEM$_{00}$ mode, the parameterized beam can then be written
in the general form by considering the first order Taylor expansion
for small modulations $(p-p_{0})/p_{0} \ll 1$
\begin{eqnarray}
u_{00}(p) \approx u_{00}(p_{0}) + (p-p_{0})
\frac{\partial{(u_{00})}}{\partial{p}}
\end{eqnarray}
where we have used $u_{ij}$ to denote the TEM$_{ij}$ Hermite Gauss
modes and $u_{ij}(p)$ denotes the same mode that experienced the
modification induced by $p$.  Specifically, a transversely
displaced and tilted beam along the x-direction is given by
\begin{eqnarray}
u_{00}(d) &=& u_{00} + d \frac{\partial{(u_{00})}}{\partial{x}} =
u_{00} + \frac{d}{w}u_{10} \\
u_{00}(\theta) &=& u_{00} + \theta
\frac{\partial{(u_{00})}}{\partial{\theta}}=u_{00} + i \frac{\pi
\theta w}{\lambda}u_{10}
\end{eqnarray}
where $d$, $\theta$, and $w$ are the displacement, tilt, and waist
diameter of the beam in the plane of observation, respectively
\cite{HSU1}.  These expressions show that small displacement
information of a Gaussian beam is encoded in the amplitude
quadrature of the co-propagating TEM$_{10}$ mode. Whilst small
tilt modulation is directly coupled to the phase quadrature of the
TEM$_{10}$ mode.

In order to extract this spatial information out of the modulated
beam, let us consider the example of homodyne detection. This
device selects the particular mode of the incoming beam which is
matched to the local oscillator transverse profile. Thus, the
detection noise mode is the one imposed by the local oscillator.
 By changing the transverse
distribution and phase of the local oscillator $\Phi_{LO}$, one
can, at will, tune the noise mode of detection, to any spatial
information of the incoming beam.  In addition, by squeezing the
noise-mode of the incoming beam, one can improve the measurement
sensitivity.  This apparatus, which we call a spatial homodyne
detector, is therefore a perfect tool for multi-mode quantum
information processing.

\begin{figure}[htbp]
\begin{center}
\includegraphics[width=8cm]{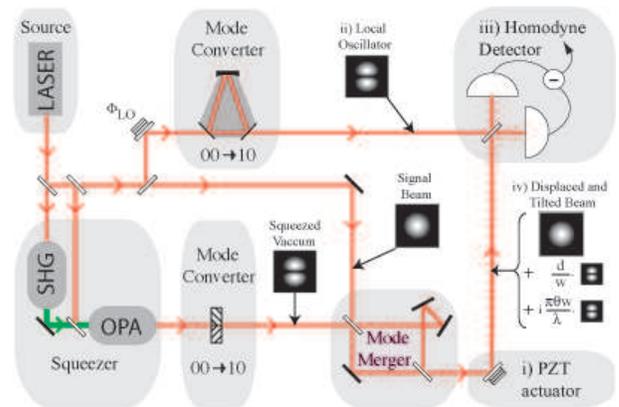}
\caption{Schematic diagram of the experiment for optimal displacement and tilt
measurements with a spatial homodyne detector.  A TEM$_{00}$ mode,
which is displaced and tilted using a PZT actuator (i), is
mode-matched to the TEM$_{10}$ local oscillator (ii) of a
balanced homodyne detector (iii).  The TEM$_{10}$ local oscillator selects
the quadratures amplitude of the TEM$_{10}$ component of the input
beam that contains the small displacement and  tilt
information of the incident TEM$_{00}$ beam (iv).}
\label{global_scheme}
\end{center}
\end{figure}

In the case of small displacement and tilt measurement, a homodyne
detector with a TEM$_{10}$ local oscillator can measure the
TEM$_{10}$ component of an incoming beam with up to $100\%$
efficiency. Hence, the detector precisely matches the displacement
and tilt conjugate observables of a TEM$_{00}$ incident beam.  A
TEM$_{10}$ spatial homodyne detector, as shown on
(Fig.~\ref{global_scheme}), is in this sense an optimal small
displacement and tilt detector.  Note that this scheme is not only
more efficient by $25\%$ than the conventional split detector to
measure a displacement \cite{HSU2}, it is also sensitive to tilt,
which is not accessible in the plane of a split detector.

As the TEM$_{10}$ mode is the noise-mode of the spatial homodyne
detector, we can experimentally improve the detection sensitivity
by filling the TEM$_{10}$ mode of the input beam with squeezed
light.
This non-classical beam is produced with an optical parametric
amplifier (OPA), that emits $3.6$ dB of vacuum squeezing in the
TEM$_{00}$ mode at $1064$ nm (note that a ring cavity - not
represented on the diagram - spatially filters the laser beam to a
pure TEM$_{00}$ mode used as the main TEM$_{00}$ mode, as well as
produces a shot noise limited beam for frequencies greater than
$1~MHz$ used to seed the OPA). A phase mask converts the vacuum
squeezed beam into a TEM$_{10}$ mode, with an efficiency of $80\%$
\cite{DEL1}, which brings the squeezing level in the TEM$_{10}$
down to $2$ dB. This vacuum squeezed TEM$_{10}$ beam is combined,
with less than $5\%$ losses, with the main bright TEM$_{00}$ beam,
by means of a modified Mach-Zehnder interferometer \cite{treps2}.
This beam interacts with a PZT actuator that induces
simultaneously displacement and tilt at RF frequencies (4~MHz).
Note that the relative amount of tilt and displacement is fixed
here by the characteristics of the actuator. This beam is analyzed
with a homodyne detector, whose TEM$_{10}$ local oscillator beam
is produced via a misaligned ring cavity resonant for the
TEM$_{10}$ mode. Note that the mode matching between these two
beams is achieved in a preliminary step by measuring a fringe
visibility of $97\%$ between the bright TEM$_{00}$ mode, and the
TEM$_{00}$ mode generated when the cavity is locked on resonance
for the TEM$_{00}$ mode instead of the TEM$_{10}$ mode.
\begin{figure}[h!]
\begin{center}
\includegraphics[width=8cm]{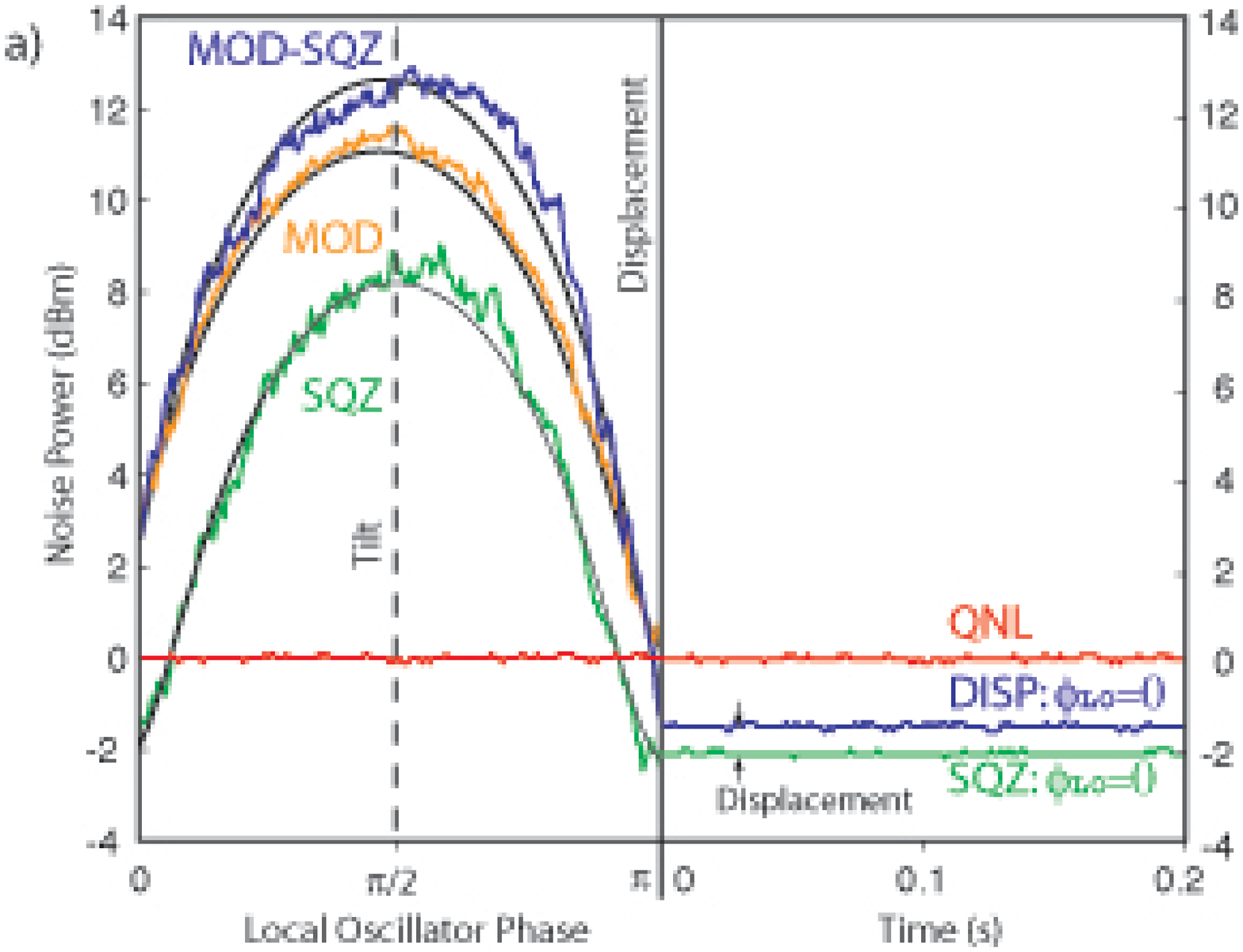}
\includegraphics[width=8cm]{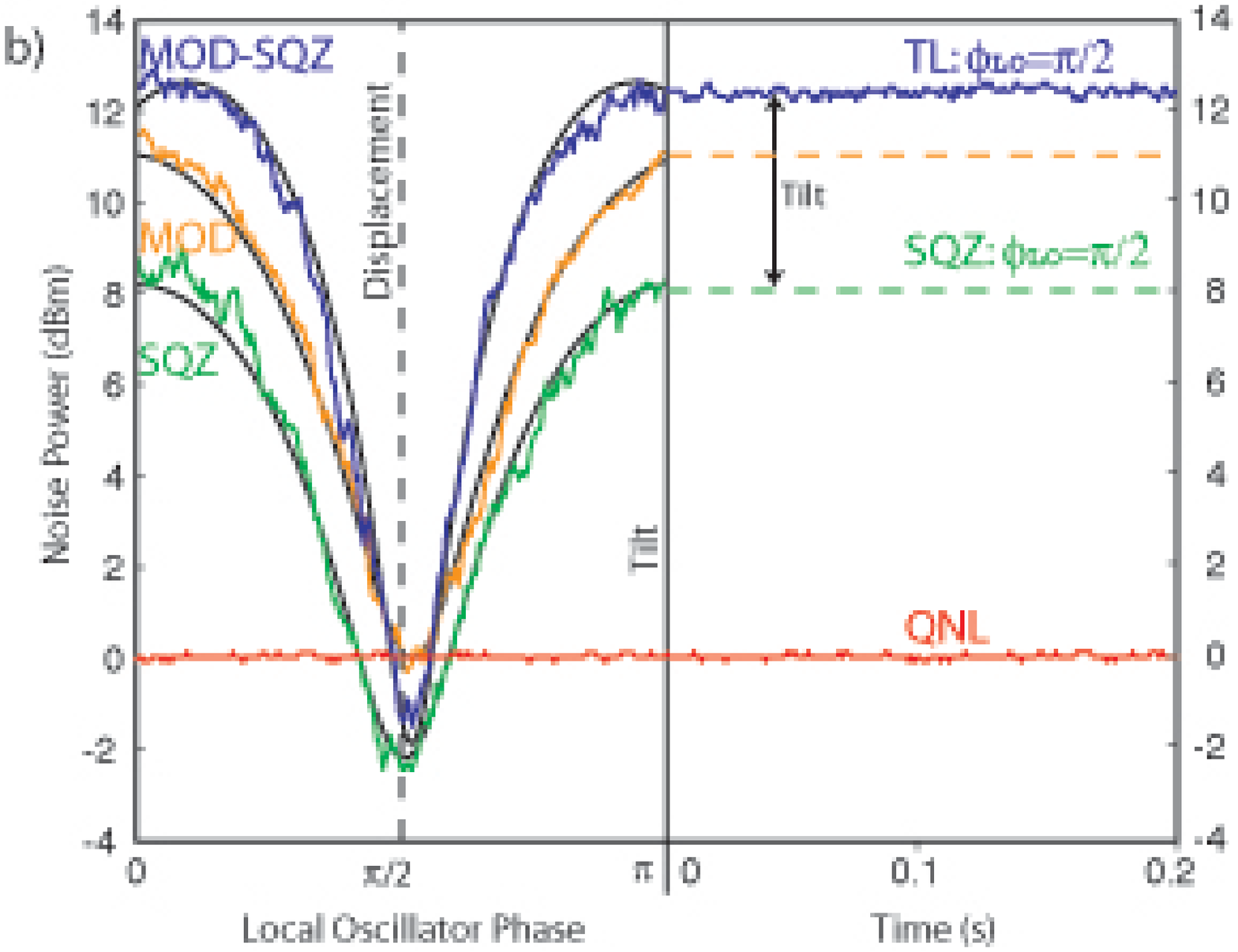}
\caption{Demonstration of sub-shot noise measurements of (a)
displacement and (b) tilt modulations using the spatial homodyne
detector.  The figures show an example where there was $90\%$ of tilt,
and only $10\%$ of displacement modulations.  Left hand side of
figures shows the scanning of local oscillator phase $\phi_{\rm LO}$
that continuously access the pure displacement (at $\phi_{\rm LO} = 0
$ and $\pi$) to pure tilt (at $\phi_{\rm LO} = \pi/2$ and $3\pi/2$)
information of the beam.  QNL: quantum noise limit.  SQZ: quadrature
noise of squeezed light with $2$dB of squeezing and $8$dB of
anti-squeezing on the TEM$_{10}$ mode.  MOD: measured modulation with
coherent light.  MOD-SQZ: measured modulation with squeezed light.
Right hand side of figures shows the corresponding locked local
oscillator phase to the (a) displacement or (b) tilt measurement.
SQZ: at $\phi_{LO}=0$ is and the squeezed noise level 2~dB below the
shot noise and at $\phi_{LO}=\pi / 2$ is 8~dB of anti-squeezing
noise.  DISP: $\phi_{LO}=0$ displacement measurement. TL: $\phi_{LO}=\pi / 2$
tilt measurement.  Displacement measurement is improved by the $2~dB$ of squeezing, while the tilt measurement is degraded by the $8~dB$ of anti-squeezing.}

\label{results}
\end{center}
\end{figure}
%

The experimental results are presented in Fig.~\ref{results}(a)
and Fig.~\ref{results}(b), when the TEM$_{10}$ local oscillator
phase is scanned and locked for displacement ($\phi_{LO}=0$) and
tilt ($\phi_{LO}=\pi / 2$) measurement.  Note that without the use
of squeezed light, the displacement modulation is masked by
quantum noise.  Improvement of the signal-to-noise ratio for
displacement measurement beyond the quantum noise limit is
achieved when the squeezed quadrature of the TEM$_{10}$ mode is in
phase with the displacement measurement quadrature (i.e. in phase
with the incoming TEM$_{00}$ mode).  Since we are dealing with
conjugated variables, improving displacement measurement degrades
the tilt measurement of the same beam, as required by the
anti-squeezing of the other quadrature.  Displacement measurement is improved by the 2~dB of squeezing, whereas the tilt measurement is degraded by the 8~dB of anti-squeezing. Theoretical curves
calculated with $2~dB$ of noise reduction, and $90~\%$ of tilt
modulation and $10~\%$ of displacement modulation - continuous
curves on Fig.~\ref{results}(a) - are in very good agreement with
experimental data.  In our experiment, we have a TEM$_{00}$ waist size of $w_{0}=106~\mu m$ in the PZT plane, a power of
$170~\mu W$, $RBW=100~kHz$ and $VBW=100~Hz$, corresponding to a Quantum
Noise Limit of  $d_{QNL}=0.6~nm$.  The measured displacement lies $1.5~dB$ below the shot noise, yielding a value of $0.4~nm$.  The ratio between displacement and tilt modulations can be inferred from the theoretical fit in figure \ref{results}, giving a measured tilt of $10^{-7}~rad$.
%


We have found and demonstrated a technique for encoding and extracting
CW quantum information on multiple co-propagating optical modes.  We
use spatial modulation as a practical technique to couple two
transverse modes and have devised a detection system whose noise mode
perfectly matches beam position and momentum variables.  We have performed quantum measurements of a pair of conjugate transverse variables, and have
verified our predictions with experiments that show the detection of a
displacement modulation below the quantum noise limit.

This work shows that in principle a large set of orthogonal multi-mode
information is accessible.  We can already simultaneously encode and
detect information in $x$ and $y$ directions \cite{TREPS3}, which
corresponds to a simultaneous use of TEM$_{10}$ and TEM$_{01}$ modes.
The possible extension to array detectors and higher order spatial
modes will be investigated.  This technique, demonstrated here in the
context of quantum imaging, leads to the feasibility of parallel
quantum information processing.

We like to thank Claude Fabre, Magnus Hsu, Warwick Bowen,
Nicolai Grosse for stimulating discussions, and Shane Grieves for
technical support.  This work was made possible by the support of
the Australian Research Council COE program.  Laboratoire Kastler
Brossel, of the Ecole Normale Superieure and University Pierre et
Marie Curie, is associated to CNRS.

\end{document}